\documentclass[preprint,showpacs,superscriptaddress,
preprintnumbers,amsmath,amssymb]{revtex4}
\usepackage{amssymb}

\usepackage{amsfonts}
\usepackage{amsmath}
\usepackage{amssymb}
\usepackage{epsfig}
\usepackage{epstopdf}
\usepackage{graphicx}%

\begin{document}
\title{Experimental investigation of the entanglement-assisted entropic uncertainty principle}

\author{Chuan-Feng Li\footnote{email: cfli@ustc.edu.cn}\footnote{These authors contributed equally to this
work.}} \affiliation{Key Laboratory of Quantum Information,
University of Science and Technology of China, CAS, Hefei, 230026,
People's Republic of China}

\author{Jin-Shi Xu$^{\dag}$}
\affiliation{Key Laboratory of Quantum Information, University of
Science and Technology of China, CAS, Hefei, 230026, People's
Republic of China}

\author{Xiao-Ye Xu}
\affiliation{Key Laboratory of Quantum Information, University of
Science and Technology of China, CAS, Hefei, 230026, People's
Republic of China}

\author{Ke Li}
\affiliation{Center for Quantum Technologies, National University of
Singapore, 2 Science Drive 3, Singapore 117542}

\author{Guang-Can Guo}
\affiliation{Key Laboratory
of Quantum Information, University of Science and Technology of
China, CAS, Hefei, 230026, People's Republic of China}
\date{\today }

\begin{abstract}
The uncertainty principle, which bounds the uncertainties involved in obtaining precise outcomes for two complementary variables defining a quantum particle, is a crucial aspect in quantum mechanics. Recently, the uncertainty principle in terms of entropy has been extended to the case involving quantum entanglement. With previously obtained quantum information for the particle of interest, the outcomes of both non-commuting observables can be predicted precisely, which greatly generalises the uncertainty relation. Here, we experimentally investigated the entanglement-assisted entropic uncertainty principle for an entirely optical setup. The uncertainty is shown to be near zero in the presence of quasi-maximal entanglement. The new uncertainty relation is further used to witness entanglement. The verified entropic uncertainty relation provides an intriguing perspective in that it implies the uncertainty principle is not only observable-dependent but is also observer-dependent.
\end{abstract}

\maketitle

In quantum mechanics, the outcomes of an observable can be predicted precisely by preparing eigenvectors corresponding to the state of the measured system. However, the ability to predict the precise outcomes of two conjugate observables for a particle is restricted by the uncertainty principle. Originally observed by Heisenberg \cite{Heisenberg27}, the uncertainty principle is best known as the Heisenberg-Robertson commutation \cite{Robertson29}
\begin{equation}
\Delta R\Delta S\geq\frac{1}{2}|\langle[R,S]\rangle|,
\end{equation}
where $\Delta R$ ($\Delta S$) represents the standard deviation of the corresponding variable $R$ ($S$). It can be seen that the bound on the right-hand side is state-dependent and can vanish even when $R$ and $S$ are non-commuting. To avoid this defect, the uncertainty relation has been re-derived in terms of an information-theoretic model of language \cite{Bialynicki75} in which the uncertainty related to the outcomes of the observable is characterized by the Shannon entropy instead of the standard deviation. The entropic uncertainty relation for any two general observables was first given by Deutsch \cite{Deutsch83}. Soon afterwards, an improved version was proposed by Kraus \cite{Kraus87} and then proved by Maassen and Uiffink
\cite{Maassen88}. The improved relation reads as follows:
\begin{equation}
H(R)+H(S)\geq\log_{2}\frac{1}{c},
\end{equation}
where $c=\max_{i,j}|\langle a_{i}|b_{j}\rangle|^{2}$ and and represents the overlap between observables $R$ and $S$, and
$|a_{i}\rangle$ ($|b_{j}\rangle$) represent the eigenvectors of the
observable $R$ ($S$).

Although we cannot obtain both the precise outcomes of two conjugate variables, even when the density matrix of the prepared state is known, the situation would be different if we invoked the effect of quantum entanglement. The possibility of violating the Heisenberg-Robertson uncertainty relation was identified early by Einstein, Podolsky, and Rosen in their famous paper, which  was originally used to challenge the correctness of quantum mechanics (EPR paradox) \cite{Einstein35}. Popper also proposed a practical experiment \cite{Popper34} to demonstrate the violation of the Heisenberg-Robertson uncertainty relation, which has since been experimentally realized \cite{Kim99}. The gedanken experiment for the EPR paradox was further exploited \cite{Reid88,Reid89}, and was experimentally demonstrated \cite{Ou92}. Currently, the violation of uncertainty relations is implemented as a signature of entanglement \cite{Hofmann03} and is used to study the continuous variable entanglement \cite{Howell04,Bowen03}.

However, the previous experimental tests were restricted to non-entropic uncertainty relations, where, crucially, the information about the initial state is purely classical. More recently, a stronger entropic uncertainty relation that uses previously determined quantum information was proved by Berta {\it et al.} \cite{Berta10}, whose equivalent form was previously conjectured by Renes and Boileau \cite{Renes09}. By initially entangling an interested particle ($A$) to another particle that acts as a quantum memory ($B$), the uncertainty associated with the outcomes of two conjugate observables can be drastically reduced to being arbitrarily small. The entropic uncertainty relation is mathematically expressed as follows \cite{Berta10}
\begin{equation}
H(R|B)+H(S|B)\geq\log_{2}\frac{1}{c}+H(A|B), \label{Unct}
\end{equation}
where $H(R|B)$ ($H(S|B)$) is the conditional von Neumann entropy
representing the uncertainty of the measurement outcomes of $R$
($S$) obtained via the information stored in $B$. $H(A|B)$ represents the conditional von Neumann entropy between $A$ and $B$. It is known that $-H(A|B)$ gives the lower bound of the
one-way distillable entanglement \cite{Devetak05}. As a result, the
lower bound of the uncertainty is essentially dependent on the entanglement
between $A$ and $B$.

In this paper, we report an experimental investigation of the new entropic uncertainty principle in a completely optical setup. This study differs from earlier related works that were mainly intended to show a violation of the classical uncertainty relation. The entropic uncertainty relation is used to witness entanglement \cite{Berta10}. We further change the complementarity of the two measured observables and verify the novel uncertainty relation (3) with the particle $B$ stored in a spin-echo based quantum memory.

We first choose to measure two Pauli observables, $R=\sigma_{x}$ and $S=\sigma_{z}$ to investigate the novel entropic uncertainty principle. The photon of interest $A$ is then prepared for entanglement with another photon $B$ via the form of different kinds of Bell diagonal states
\begin{equation}
\rho_{1}=x|\Phi^{+}\rangle\langle\Phi^{+}|+(1-x)|\Psi^{-}\rangle\langle\Psi^{-}|,
\label{equ:Bell1}
\end{equation}
and
\begin{equation}
\rho_{2}=x|\Phi^{-}\rangle\langle\Phi^{-}|+(1-x)|\Psi^{-}\rangle\langle\Psi^{-}|,\label{equ:Bell2}
\end{equation}
where
$|\Phi^{\pm}\rangle=\frac{1}{\sqrt{2}}(|00\rangle\pm|11\rangle)$ and
$|\Psi^{-}\rangle=\frac{1}{\sqrt{2}}(|01\rangle-|10\rangle)$ are the
Bell states. $x$ represents the corresponding ratio between these
two components in $\rho_{1}$ and $\rho_{2}$ (the calculation of corresponding conditional entropies is given in Methods).

To use the entropic uncertainty relation (\ref{Unct}) to witness
entanglement, we follow the same procedure using observables $R=\sigma_{x}$
($S=\sigma_{z}$) on both particles $A$ and $B$. The variable $d_{R}$
represents the probability that the outcomes of $R$ on $A$ and $R$
on $B$ are different, and $d_{S}$ represents the probability that the
outcomes of $S$ on $A$ and $S$ on $B$ are different. According to the
Fano's inequality relation \cite{Fano61}
\begin{equation}
H(R|B)+H(S|B)\leq h(d_{R})+h(d_{S}), \label{EWIT}
\end{equation}
where $h(d_{R})=-d_{R}\log_{2}d_{R}-(1-d_{R})\log_{2}(1-d_{R})$. As
a result, when $h(d_{R})+h(d_{S})-1<0$, $H(A|B)<0$ according to the
inequality (\ref{Unct}), which indicates the entanglement between $A$
and $B$.

\begin{figure}[tbph]
\begin{center}
\includegraphics [width= 6.0in]{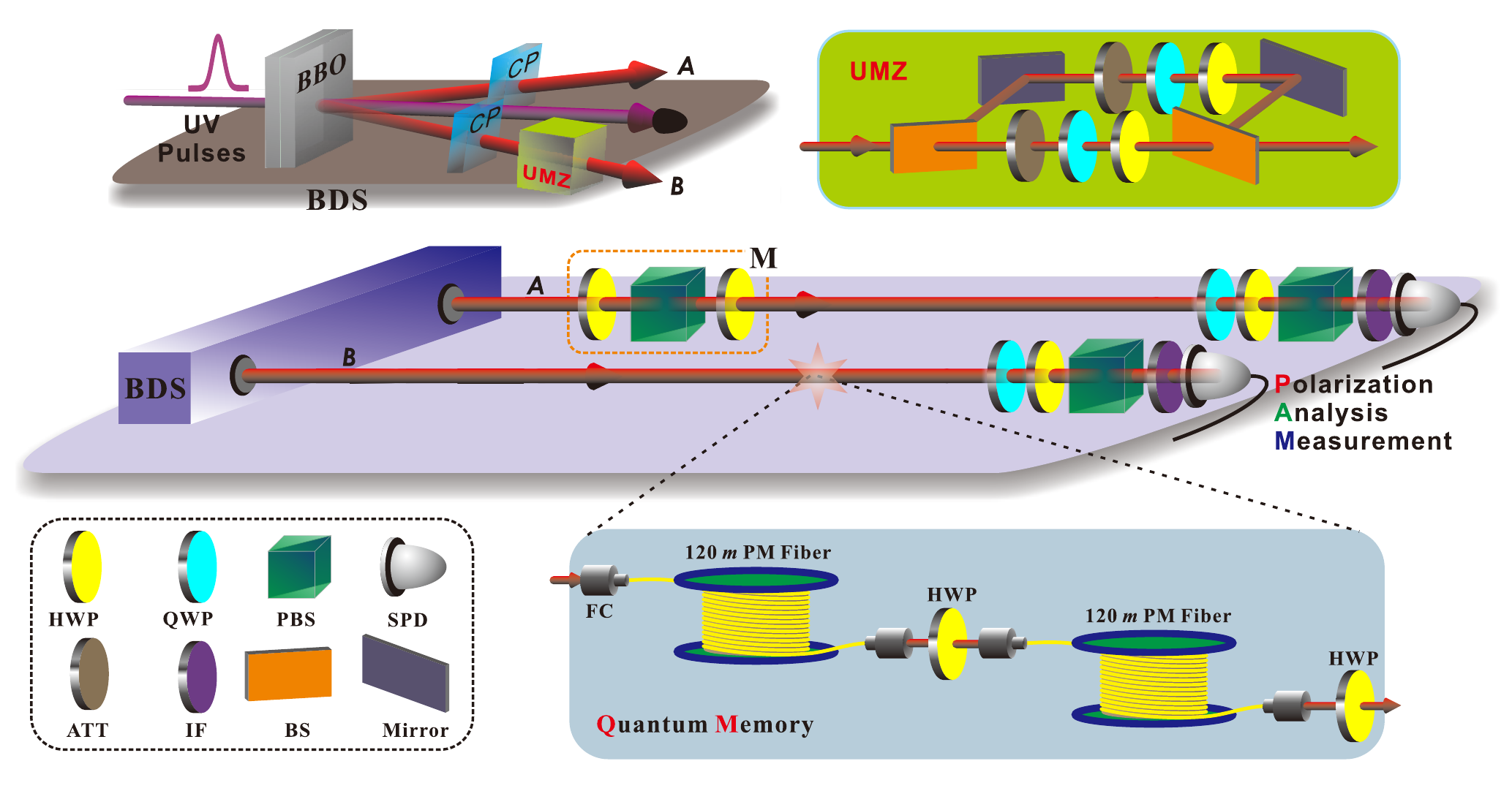}
\end{center}
\caption{(Color online). Experimental setup. Ultraviolet (UV) pulses pass through two type-I $\beta$-barium borate (BBO) crystals to produce polarization-entangled photon pairs, that are emitted into modes $A$ and $B$. Quartz plates (CP) are used to compensate the birefringence of the BBO crystals. The photon in mode $B$ further passes through an unbalanced Mach-Zenhder interference (UMZ) setup to prepare the required Bell diagonal states (BDS). The attenuators (ATT) are used to control the ratio between different components in the BDS. Quarter-wave plates (QWP) and half-wave plates (HWP) are employed to prepare the exact forms of the BDS. The dashed pane $M$ containing two HWPs and a polarization beam splitter (PBS) is used to measure $R$ and $S$ on the photon $A$. With the optic axes of the two HWPs set to $\theta/2$ and $\theta/2-45^{\circ}$, $M$ projects the corresponding state of photon $A$ onto the two eigenvectors $\cos\theta|H\rangle+\sin\theta|V\rangle$ and $\sin\theta|H\rangle-\cos\theta|V\rangle$. The quantum memory, which consists of two polarization maintaining fibers (PM Fibers) with 120 m length and two HWPs (FC represents the fiber coupler), is performed on the photon in mode $B$ depending on the specific case. The polarization analysis measurement device containing a QWP, HWP and PBS in each arm is used to perform observable measurements on both photons and the tomographic measurement. Both photons are then detected by single-photon detectors (SPDs) equipped with 3 nm interference filters (IFs). When the quantum memory is performed on mode $B$, the detected signal in mode $A$ is delayed by about 1.2 $\mu$s to coincide with that in mode $B$ in the coincidence counting circuit (not shown).}
\label{fig:setup}
\end{figure}

In our experiment, the polarizations of photons are encoded as information carriers. We set the horizontal polarization state ($|H\rangle$) as $|0\rangle$ and the vertical polarization state ($|V\rangle$) as $|1\rangle$. Figure 1 shows the experimental setup. Ultraviolet (UV) pulses with a 76 MHz repetition rate (wavelength centres at 400 nm) are focused on two type-I  $\beta$-barium borate (BBO) crystals to generate polarization-entangled photon pairs \cite{Kwiat99}, which are emitted into modes $A$ and $B$ (for simplicity, we just refer to photons $A$ and $B$). After compensating the birefringence with quartz plates (CP), the maximally entangled state $|\Phi^{+}\rangle=1/\sqrt{2}(|HH\rangle+|VV\rangle)$ is prepared with high visibility \cite{Xu06}. In order to prepare different kinds of Bell diagonal states (BDS), photon $B$ further passes through an unbalanced Mach-Zenhder interference (UMZ) setup. The time difference between the short and long paths of the UMZ is about 1.5 ns, which is smaller than the coincidence window. By tracing over the path information in the UMZ \cite{Aiello07}, the BDS described by equations (4) and (5) can be produced. The density matrix of the initial BDS is characterized by the quantum state tomography process \cite{James01}, in which $H(A|B)$ can be calculated. In order to measure $H(R|B)$ and  $H(S|B)$, the measurement apparatus $M$ containing two HWPs and a polarization beam splitter (PBS) is performed on the photon $A$. After passing through $M$, photon $A$ is sent to the polarization analysis measurement device together with photon $B$ for quantum state tomography. The spin-echo based quantum memory, consisting of two polarization maintaining fibers (PM Fibers) each of 120 m length and two HWPs with the angles set at $45^{\circ}$, is performed on mode $B$ depending on the specific case. The polarization analysis measurement setup containing QWPs, HWPs and PBSs can be used to perform corresponding observable measurements on both photons and the tomographic measurement. These two photons are then detected by two single photon detectors (SPDs) equipped with 3 nm interference filters (IFs), in which the measured quantities are based on coincident counts.

\begin{figure}[tbph]
\begin{center}
\includegraphics [width= 3.0in]{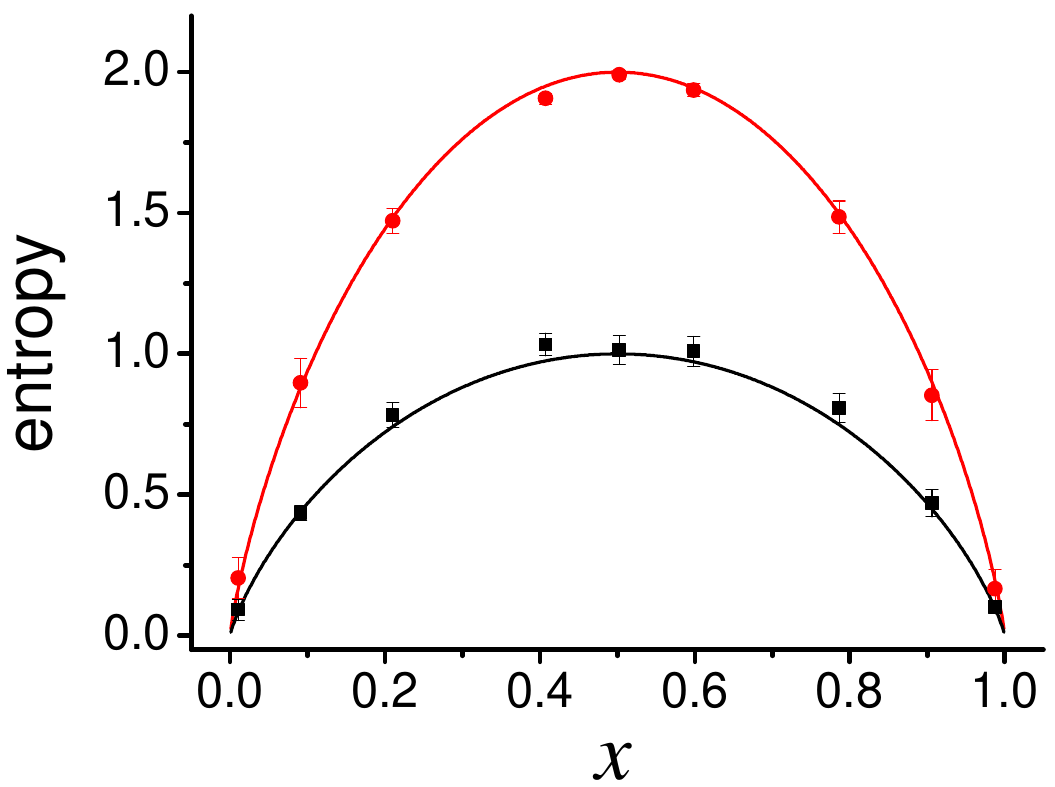}
\end{center}
\caption{(Color online). Experimental results of the conditional
entropies with the input state of equation (\ref{equ:Bell1}). The
$x$ axis represents the amount of $|\Phi^{+}\rangle$ in the state
(\ref{equ:Bell1}). Red dots represent the experimental results of
$H(\sigma_{x}|B)+H(\sigma_{z}|B)$ and black squares denote the
results of $1+H(A|B)$. The red and black solid lines are the
corresponding theoretical predictions, respectively. The state at the point of $x=0.5$ is the maximally mixed state without entanglement, where $H(\sigma_{x}|B)+H(\sigma_{z}|B)$ becomes maximal. At the points near $x=0$ and $x=1$ where photon $A$ is quasi-maximally entangled to $B$, the lower bound of $1+H(A|B)$ is near zero, and the uncertainty of $H(\sigma_{x}|B)+H(\sigma_{z}|B)$ is close to this value within the error bars. Error bars
represent the corresponding standard deviations.}
\label{fig:entropy_1}
\end{figure}

Fig. \ref{fig:entropy_1} shows the experimental results of the
uncertainties when measuring the outcomes of $\sigma_{x}$ and
$\sigma_{z}$ on the photon $A$, which is entangled with another
photon $B$ in the form of Eq. (\ref{equ:Bell1}). Red dots and black squares represent the experimental
results of $H(\sigma_{x}|B)+H(\sigma_{z}|B)$ and $1+H(A|B)$,
with the red and black solid lines representing the corresponding theoretical predictions, respectively. It is clear that $1+H(A|B)$ provides a lower bound of uncertainties when obtaining the outcomes of both $\sigma_{x}$ and $\sigma_{z}$ and the experimental results agree well with the theoretical predictions within error bars.

\begin{figure}[tbph]
\begin{center}
\includegraphics [width= 3.0in]{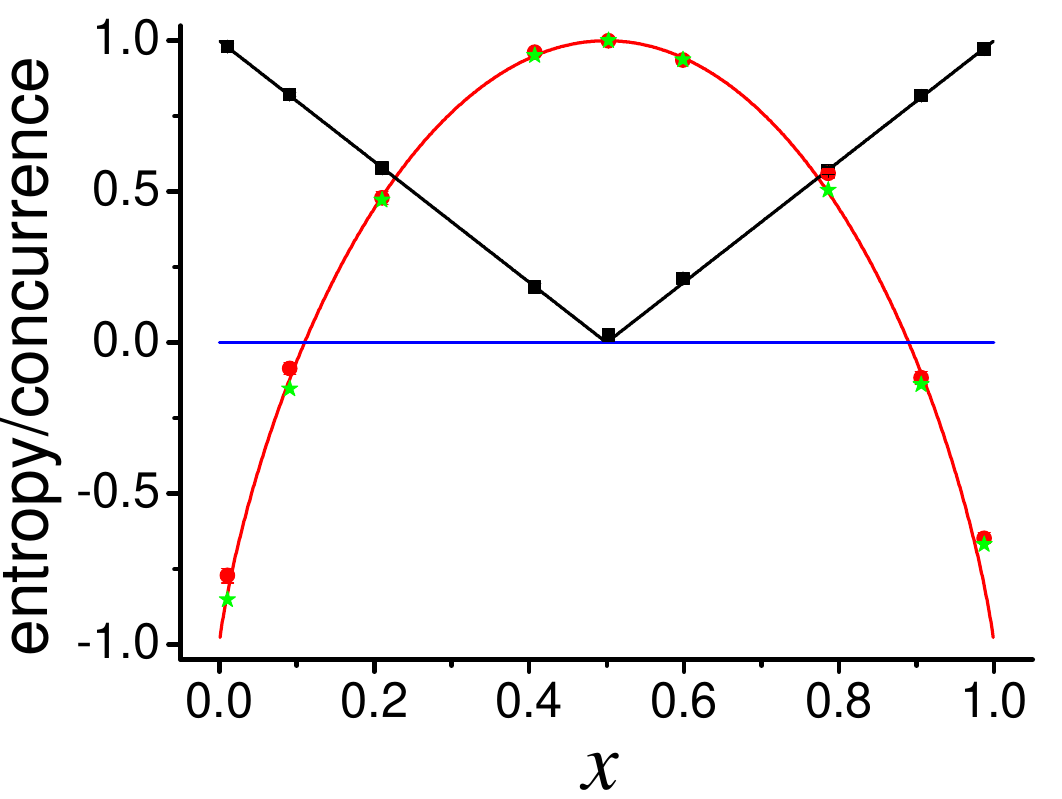}
\end{center}
\caption{(Color online). Experimental results of the entanglement
witness with the input state of equation (\ref{equ:Bell1}). The $x$
axis represents the amount of $|\Phi^{+}\rangle$ in the state
(\ref{equ:Bell1}). Red dots represent the experimental results of
$h(d_{R})+h(d_{S})-1$ with the red solid line representing the
theoretical prediction. The blue solid line represents the constant
of zero. The green stars are the calculated values of
$h(d_{R})+h(d_{S})-1$ from the measured density matrix. Black
squares denote the results of concurrence with the black solid line
representing the theoretical prediction. The concurrence is always larger than 0 except for the state at $x=0.5$, which represents the separated state. Error bars represent the corresponding standard deviations }
\label{fig:entanglement_1}
\end{figure}

Next, we use the entropic uncertainty relation of inequality
(\ref{Unct}) to witness entanglement. Fig. \ref{fig:entanglement_1} shows the experimental
results. The red dots represent the experimental results of $h(d_{\sigma_{x}})+h(d_{\sigma_{z}})-1$, and the red solid line represents the corresponding theoretical prediction (see Methods for its calculation). The cases with $d(\sigma_{x})+h(d_{\sigma_{z}})-1<0$ indicate a one-way distillable entanglement between $A$ and $B$ \cite{Devetak05}. The blue solid line represents a constant of zero. The green stars denote the theoretically calculated value of $h(d_{\sigma_{x}})+h(d_{\sigma_{z}})-1$ from the experimentally measured density matrix of $\rho_{1}$, which agrees with the experimental results. The entanglement between $A$ and $B$ is further measured by the concurrence \cite{Wootters98} represented by the black squares, and the black solid line represents the theoretical prediction (see Methods). We can see from fig. 3 that the value of $h(d_{\sigma_{x}})+h(d_{\sigma_{z}})-1$ witnesses lower bounds of entanglement shared between $A$ and $B$. The concurrence is calculated from the reconstructed density matrix requires quantum state tomography with 9 measurement settings, while the approach using the uncertainty relation to witness entanglement requires only 2 measurement settings. Thus, this novel uncertainty relation would find practical use in the area of quantum engineering.

\begin{figure}[tbph]
\begin{center}
\includegraphics [width= 3.0in]{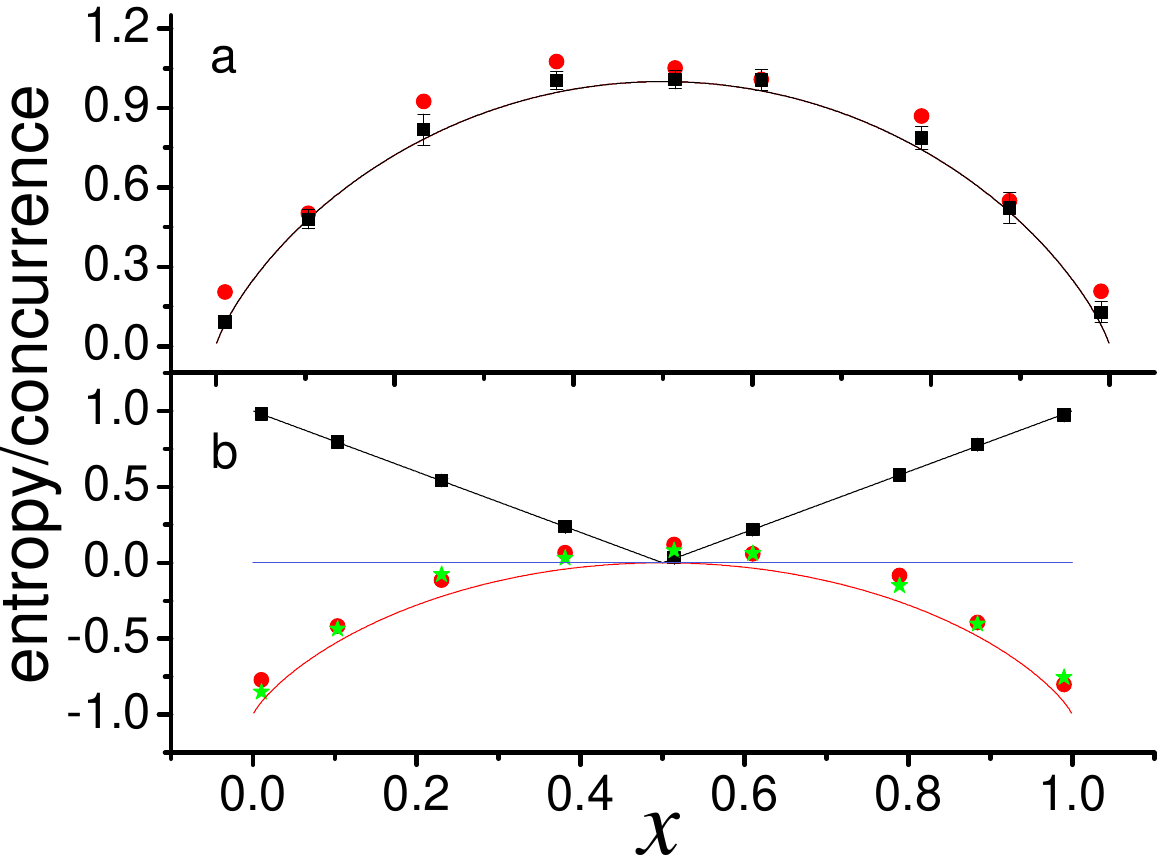}
\end{center}
\caption{(Color online). Experimental results with the input state
(\ref{equ:Bell2}). The $x$ represents the amount of
$|\Phi^{-}\rangle$ in the state (\ref{equ:Bell2}). (a) Experimental
results of the conditional entropies. Red dots represent the values
of $H(\sigma_{x}|B)+H(\sigma_{z}|B)$ and black squares represent
$1+H(A|B)$. The red and black solid lines represent the
corresponding theoretical predictions, which overlap completely and
only the solid line can be seen. (b) Experimental results of the
entanglement witness. Red dots represent the results of
$h(d_{\sigma_{x}})+h(d_{\sigma_{z}})-1$ with the red solid line
representing the theoretical prediction. The green stars are the
calculated values of $h(d_{\sigma_{x}})+h(d_{\sigma_{z}})-1$ from
the measured density matrix. The blue solid line represents the constant
of zero. Black squares denote the results of
concurrence with the black solid line representing the theoretical
prediction.} \label{fig:combine}
\end{figure}

We further consider another case in which the prepared state is the
Bell diagonal state (\ref{equ:Bell2}). Fig. \ref{fig:combine}(a)
shows the experimental results of the conditional entropies. Red
dots represent the values of $H(\sigma_{x}|B)+H(\sigma_{z}|B)$ and
black squares represent $1+H(A|B)$. They are equal to each other
within the error bars. The red and black solid lines represent the
corresponding theoretical predictions, which overlap completely and
only the black solid line can be seen. Compared
with the case in fig. \ref{fig:entropy_1}, we find that the
conditional entropy of $H(\sigma_{x}|B)+H(\sigma_{z}|B)$ is not only
dependent on the entanglement between $A$ and $B$, but also
dependent on the exact form of the entangled state. However,
$H(\sigma_{x}|B)+H(\sigma_{z}|B)$ can be always close to zero with
arbitrary precision in the present of maximally entangled states.
The results of applying the inequality (\ref{Unct}) to witness
entanglement is shown in fig. \ref{fig:combine}(b). Red dots
represent the experimental results of
$h(d_{\sigma_{x}})+h(d_{\sigma_{z}})-1$ with the red solid line
representing the theoretical prediction. Green stars are the
calculated values of $h(d_{\sigma_{x}})+h(d_{\sigma_{z}})-1$ from
the measured density matrix, which agree well with the red dots. The
black squares represent the experimental results of concurrence with
the black solid line representing the theoretical prediction. The
concurrence and the value of $h(d_{\sigma_{x}})+h(d_{\sigma_{z}})-1$
are coincident in witnessing entanglement for when the concurrence
is larger than zero, the value of
$h(d_{\sigma_{x}})+h(d_{\sigma_{z}})-1$ is smaller than zero. They
overlap at the point of $x=0.5$, which represents the separated
state. The blue solid line represents the constant
of zero. Combined with the analysis of fig. \ref{fig:entanglement_1},
we can find that the value of
$h(d_{\sigma_{x}})+h(d_{\sigma_{z}})-1$ is also dependent on the
exact form of the prepared entangled state.

\begin{figure}[tbph]
\begin{center}
\includegraphics [width= 3.0in]{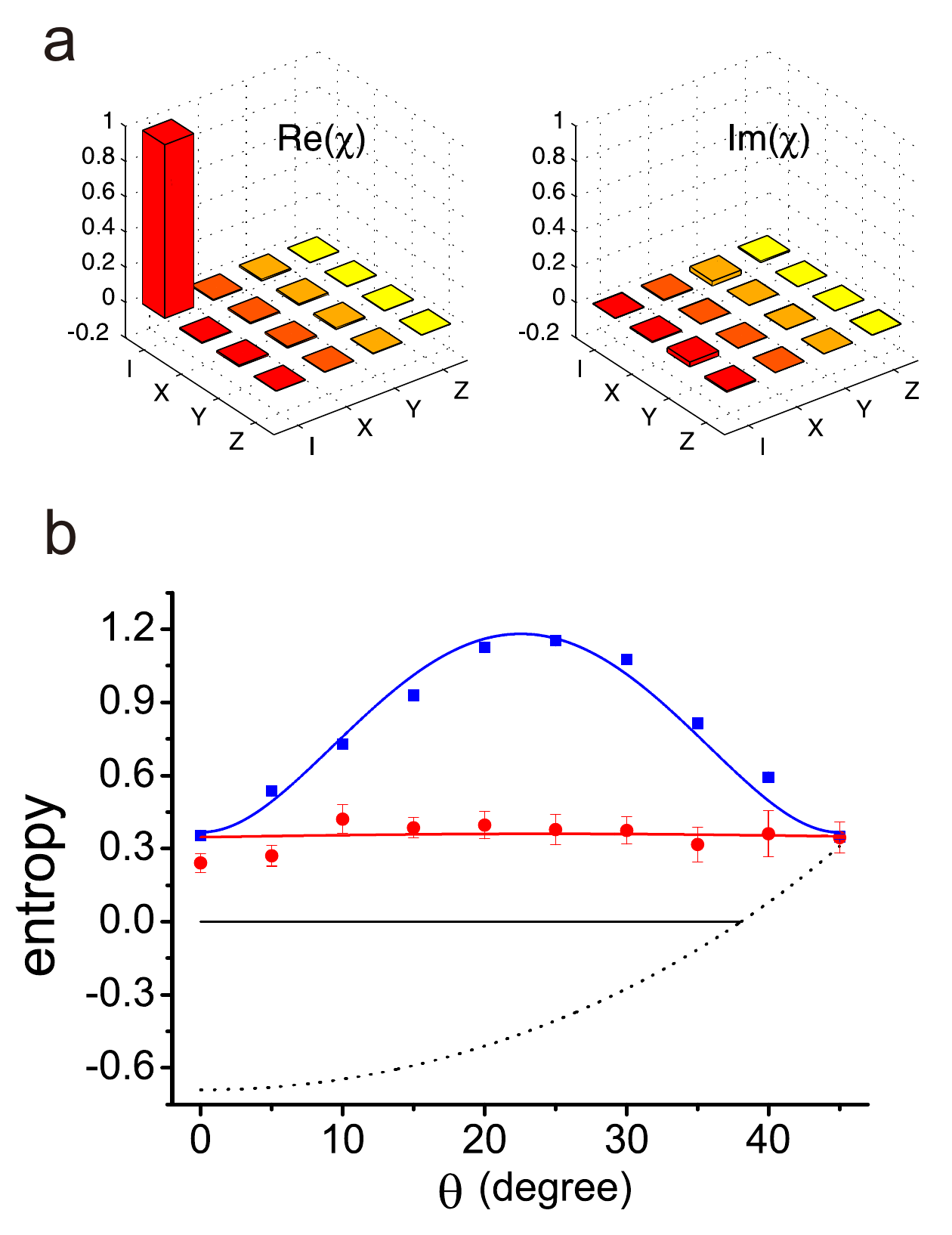}
\end{center}
\caption{(Color online). Experimental results for the density matrix $\chi$ of the spin-echo based quantum memory and the entropies as a function of the angle $\theta$. (a) $Re(\chi)$ represents the real part of $\chi$ and $Im(\chi)$ represents the imaginary part of $\chi$. (b) The initial input state of $AB$ system with $B$ passing through the quantum memory was prepared as a quasi-maximal entangled state with the form close to $1/\sqrt{2}(|HH\rangle-|VV\rangle)$, and the concurrence equals approximately 0.921 with a relative high entropy value for $H(A|B)$ of about -0.692. The red dots and blue squares represent the experimental results of $H(R|B)+H(S|B)$ and  $H(R|R)+H(S|S)$, respectively. The red and blue solid lines represent the corresponding theoretical predictions, which agree with the experimental results. Error bars represent the standard deviations (error bars of $H(R|R)+H(S|S)$ are smaller than the corresponding symbols). The black dotted line represents the theoretical prediction of the lower bound of the uncertainty relation (3). When the lower bound is smaller than zero, it is set to zero (black solid line).} \label{fig:method}
\end{figure}

We then consider the case storing photon $B$ in a spin-echo based quantum memory. Fig. 5(a) shows the real ($Re$) and imaginary ($Im$) parts of the density matrix $\chi$ characterizing the operation of the quantum memory (a detailed description of it is contained in the Methods). The operation of the optical delay closes to the identity, which serves as a high-quality quantum memory with a fidelity of about $98.3\%$. Fig. 5(b) shows experimental results obtained for the uncertainties as a function of the angle $\theta$. We use two methods to estimate the uncertainty (see Methods). The red dots and blue squares represent the experimental results of $H(R|B)+H(S|B)$ and $H(R|R)+H(S|S)$, respectively. The uncertainty estimated via direct measurements of both $A$ and $B$ ($H(R|R)+H(S|S)$) is never less than the uncertainty estimated via the process of quantum state tomography ($H(R|B)+H(S|B)$), which provides an upper bound of the novel uncertainty relation (3). The lower bound of the novel uncertainty relation $\log_{2}(\frac{1}{c(\theta)})+H(A|B)$ is less than zero when $\theta<38^{\circ}$, requiring that it be set to be zero. Error bars represent the standard deviations.

In conclusion, we have experimentally investigated the entropic uncertainty relation with the assistance of entanglement. In addition, this study verifies the application of the entropic uncertainty relation to witness the distillable entanglement assisted by one-way classical communication from $A$ to $B$. Although the value of $h(d_{\sigma_{x}})+h(d_{\sigma_{z}})-1$ is dependent on the exact form of entangled states, it can be obtained by a few separate measurements on each of the entangled particles \cite{Berta10}, which shows its ease of accessibility. The method used to estimate uncertainties by directly performing measurements on both photons has the practical application in verifying the security of quantum key distribution \cite{Berta10}. Our results not only violate the previous classical uncertainty relation but also confirm the novel one proposed by Berta {\it et al.} \cite{Berta10}. The verified entropic uncertainty principle implies that the uncertainty principle is not only observable-dependent but is also observer-dependent \cite{Winter10}, providing a particularly intriguing perspective. While preparing our manuscript for submission, we noted that another relevant experimental work was performed independently by Prevedel {\it et al.} \cite{Prevedel11}.

\section*{Methods}
{\bf Conditional entropies for $\rho_{1}$ and $\rho_{2}$.} If the two observables are chose to be $R=\sigma_{x}$ and $S=\sigma_{z}$, the eigenvectors of $R$ are $|D\rangle=1/\sqrt{2}(|0\rangle+|1\rangle)$, and $|J\rangle=1/\sqrt{2}(|0\rangle-|1\rangle)$, and the eigenvectors of $S$ are $|0\rangle$ and $|1\rangle$.  As a result, the maximal complementarity ($c$) between $R$ and $S$ is $1/2$ and $\log_{2}(\frac{1}{c})=1$. For the initial input state $\rho_{1}$, the conditional von Neumann entropy on the left-hand side of the inequality (3) is calculated to be $H(R|B)+H(S|B)=H(\sigma_{\sigma_{x}})+H(\sigma_{z}|B)=-2x\log_{2}x-2(1-x)\log_{2}(1-x)$ and the right-hand side is calculated as $\log_{2}\frac{1}{c}+H(A|B)=-x\log_{2}x-(1-x)\log_{2}(1-x)$. As a result, $\log_{2}\frac{1}{c}+H(A|B)$ gives the lower bound of $H(\sigma_{x}|B)+H(\sigma_{z}|B)$ ($0\leq x\leq1$). At the points of $x=0$ and $x=1$, i.e., $\rho_{1}$ represents the maximally entangled state, for which the left-hand term and the right-hand term both equal 0. For the input state $\rho_{2}$, the conditional entropies are calculated as $H(\sigma_{x}|B)+H(\sigma_{z}|B)=\log_{2}\frac{1}{c}+H(A|B)=-x\log_{2}x-(1-x)\log_{2}(1-x)$. Therefore, the right-hand term gives the exact bound in the uncertainty relation (3).

{\bf Calculation of novel entanglement witness.} In order to obtain the values of $h(d_{R})+h(d_{S})-1$, the observable measurements ($R=\sigma_{x}$ and $S=\sigma_{z}$) on both photons are directly preformed by the polarization analysis measurement setup (fig. 1). The probabilities of obtaining the different outcomes of $\sigma_{x}$ ($\sigma_{z}$) on $A$ and $B$ are calculated as $d_{\sigma_{x}}=(N_{DJ}+N_{JD})/(N_{DD}+N_{DJ}+N_{JD}+N_{JJ})$ ($d_{\sigma_{z}}=(N_{HV}+N_{VH})/(N_{HH}+N_{HV}+N_{VH}+N_{VV})$), where $N_{ij}$ represents the coincident counts when the photon state of $A$ is projected onto $|i\rangle$ and $B$ is projected onto $|j\rangle$ ($|i\rangle,|j\rangle\in\{|D\rangle,|J\rangle,|H\rangle,|V\rangle\}$).

{\bf Concurrence.} For a two-qubit state $\rho$, the concurrence \cite{Wootters98} is given by
$C=\max\{0,\Gamma\}$,
where $\Gamma=\sqrt{\lambda_{1}}-\sqrt{\lambda_{2}}-\sqrt{\lambda_{3}}-\sqrt{\lambda_{4}}$, and the quantities $\lambda_{j}$ are the eigenvalues in decreasing order of the matrix $\rho(\sigma_{y}\otimes\sigma_{y})\rho^{*}(\sigma_{y}\otimes\sigma_{y})$ with $\sigma_{y}$ denoting the second Pauli matrix. The variable $\rho^{*}$ corresponds to the complex conjugate of $\rho$ in the canonical basis $\{|00\rangle,|01\rangle,|10\rangle,|11\rangle\}$.

{\bf Quantum memory.} In our experiment, the quantum memory is constructed via two polarization maintaining fibers (PM Fibers) each of 120 m length and two half-wave plates with the angles set at $45^{\circ}$, as shown in fig. 1. Both PM fibers are set at the same preference basis $\{|H\rangle,|V\rangle\}$. Consider a photon with the polarization state $\alpha|H\rangle+\beta|V\rangle$ ($\alpha$ and $\beta$ are the two complex coefficients of the corresponding polarization states $|H\rangle$ and $|V\rangle$) passing through one of the fibers. Due to the different indices of refraction of horizontal and vertical polarization in the PM fiber, different phases impose on the corresponding polarization states, which can be written as $\alpha e^{i\phi_{H}}|H\rangle+\beta e^{i\phi_{V}}|V\rangle$ for simplicity. A half-wave plate is then implemented by exchanging $|H\rangle$ and $|V\rangle$. After the photon passes the same second PM fiber, the state becomes $e^{i(\phi_{H}+\phi_{V})}(\alpha|V\rangle+\beta|H\rangle)$ and the coherence of the state is recovered. We then apply another half-wave plate to exchange $|H\rangle$ and $|V\rangle$, and the state becomes the initial form. This process is similar to the phenomenon of spin echo in nuclear magnetic resonance, and the photon is stored in the PM fibers for about 1.2 $\mu$s. Therefore, this system may serve as a spin-echo based quantum memory.

We then characterize the spin-echo based quantum memory by using the quantum process tomography \cite{Chuang97}. Its operator can be expressed on the basis of $\hat{E_{m}}$   and written as $\varepsilon=\Sigma_{mn}\chi_{mn}\hat{E_{m}}\rho\hat{E_{n}}^{\dag}$. The basis of $\hat{E_{m}}$ we chose is $\{I,X,Y,Z\}$, where $I$ represents the identical operation and $X$, $Y$ and $Z$ represent the three Pauli operators, respectively. The matrix $\chi$ completely and uniquely describes the process $\varepsilon$ and can be reconstructed by experimental tomographic measurements. In the experiment, the physical matrix $\chi$ is estimated by the maximum-likelihood procedure \cite{OBrien04}, which is represented in fig. 5(a). It is closed to the identity and the fidelity of the experimental result is about $98.3\%$, which is calculated by $(Tr(\sqrt{\sqrt{\chi}\chi_{ideal}\sqrt{\chi}}))$ with $\chi_{ideal}=I$. As a result, the spin-echo based optical delay acts as a high-quality quantum memory.

{\bf Estimation of uncertainties with quantum memory.} In the experiment employing quantum memory, we change the complementarity of the two observables to be measured. The operator $S$ is chosen to be $\sigma_{z}$ with the eigenvectors $|H\rangle$ and $|V\rangle$, while the other operator $R$ is chosen to be in the X-Z plane with the eigenvectors $\cos\theta|H\rangle+\sin\theta|V\rangle$ and $\sin\theta|H\rangle-\cos\theta|V\rangle$. As a result, the complementarity of these observables becomes $c(\theta)=-\log_{2}\max[\cos\theta^{2},\sin\theta^{2}]$. We use two methods to estimate the uncertainty. The first is based on the quantum state tomography, which is given by the conditional von Neumann entropy $H(R|B)+H(S|B)$. The other quantity, directly estimated by the coincidence counts used for the same measurements on both $A$ and $B$, is represented by $H(R|R)+H(S|S)$. For example, $H(\sigma_{z}|\sigma_{z})=-\Sigma_{i,j=\{H,V\}}(N_{ij}/N))\log_{2}(N_{ij}/N)+\Sigma_{k=\{1,2\}}(N_{k}/N)\log_{2}(N_{k}/N)$, in which $N$ represents the total coincidence counts and $N_{1}=N_{HH}+N_{VH}$ ($N_{2}=N_{HV}+N_{VV}$ ) represents the counts when the state of photon $B$ is projected onto $|H\rangle$ ($|V\rangle$) by tracing out the photon $A$. Because  $H(R|R)+H(S|S)\geq H(R|B)+H(S|B)$, $H(R|R)+H(S|S)$ provides an upper bound of the novel uncertainty relation (3).

{\bf Error estimation.} In our experiment, the pump power is about 100 mW, and the total coincident counts are about 6000 in 30 s. The statistical variation of each count is considered according to the Poisson distribution, and the error bars are estimated from the standard deviations of the values calculated by the Monte Carlo method \cite{Altepeter05}.

\section*{Acknowledgements}

This work was supported by the National Basic Research Program of China (Grants No. 2011CB9212000), National Natural Science Foundation of China (Grant Nos. 11004185, 60921091, 10874162), and the China Postdoctoral Science Foundation (Grant No. 20100470836). The CQT is funded by the Singapore MoE and the NRF as part of the Research Centres of Excellence programme.

\end{document}